\documentclass[aps,prd,superscriptaddress,twocolumn,nofootinbib]{revtex4-1}
\usepackage[utf8]{inputenc}
\usepackage{dsfont}
\usepackage{amsfonts}
\usepackage{amsmath}
\allowdisplaybreaks[4]        
\usepackage{amssymb}
\usepackage{euscript}     
\usepackage{braket}
\usepackage{tikz}
\usepackage{starfont}
\usepackage{color,soul}         
\usepackage{tensor}        
\usepackage{amsthm}
\usepackage{graphicx}
\usepackage{slashed}
\usepackage{leftidx}
\usepackage{subfigure}
\usepackage{bbm}
\usepackage{comment}
\definecolor{outerspace}{rgb}{0.25, 0.29, 0.3}
\definecolor{scarlet}{rgb}{1.0, 0.13, 0.0}
\usepackage[header,title,page,titletoc]{appendix}  
\definecolor{princetonorange}{rgb}{1.0, 0.56, 0.0}
\definecolor{WildStrawberry}{rgb}{1.0, 0.26, 0.64}
\definecolor{rossocorsa}{rgb}{0.83, 0.0, 0.0}
\definecolor{navyblue}{rgb}{0.0, 0.0, 0.5}
\usepackage{float}
\usepackage[paper=letterpaper,margin=1in]{geometry}
\usepackage{mathtools}

\newcommand{\req}[1]{(\ref{#1})} 

\newcommand{\bea}{\begin{eqnarray}}

\newcommand{\eea}{\end{eqnarray}}
\newcommand{\ba}{\begin{eqnarray}}
\newcommand{\ea}{\end{eqnarray}}

\newcommand{\be}{\begin{equation}}
\newcommand{\ee}{\end{equation} }
\newcommand{\beqa}{\begin{eqnarray}}
\newcommand{\eeqa}{\end{eqnarray}}
\newcommand{\beqar}{\begin{eqnarray*}}
\newcommand{\eeqar}{\end{eqnarray*}}

\renewcommand{\req}[1]{(\ref{#1})}

\newcommand{\drm}{\mathrm{d}}

\renewcommand{\c}{$c$}

\newcommand{\grad}{\nabla}

\newcommand{\tcb}{\textcolor{blue}}

\makeatletter
\newcommand{\dal}{\mathop{\mathpalette\dal@\relax}}
\newcommand{\dal@}[2]{%
  \begingroup
  \sbox\z@{$\m@th#1\square$}%
  \dimen0=\fontdimen8
    \ifx#1\displaystyle\textfont\else
    \ifx#1\textstyle\textfont\else
    \ifx#1\scriptstyle\scriptfont\else
    \scriptscriptfont\fi\fi\fi3
  \makebox[\wd\z@]{%
    \hbox to \ht\z@{%
      \vrule width \dimen0
      \kern-\dimen0
      \vbox to \ht\z@{
        \hrule height \dimen0 width \ht\z@
        \vss
        \hrule height 2\dimen0
      }%
      \kern-2.5\dimen0
      \vrule width 2.5\dimen0
    }%
  }%
  \endgroup
}
\makeatother

\usepackage{hyperref}
\hypersetup{
    colorlinks,
    citecolor=blue,
    filecolor=navyblue,
    linkcolor=navyblue,
    urlcolor=navyblue
}

\begin{document}

\title{Scalar fields from nonlinear sigma models on black hole spacetimes}
\author{Ludovico Machet}
\email{ludovico.machet@kuleuven.be}
\affiliation{Instituut voor Theoretische Fysica, KU Leuven. Celestijnenlaan 200D, B-3001 Leuven, Belgium. }
\affiliation{Leuven Gravity Institute, KU Leuven. Celestijnenlaan 200D, B-3001 Leuven, Belgium. }
\affiliation{Physique Théorique et Mathématique, ULB, Boulevard du Triomphe, B-1050 Bruxelles, Belgium}

\author{Llibert Arest\'e Sal\'o}
\email{llibert.arestesalo@kuleuven.be}
\affiliation{Instituut voor Theoretische Fysica, KU Leuven. Celestijnenlaan 200D, B-3001 Leuven, Belgium. }
\affiliation{Leuven Gravity Institute, KU Leuven. Celestijnenlaan 200D, B-3001 Leuven, Belgium. }

\begin{abstract}
Scalar fields with non-trivial kinetic term derived from a nonlinear sigma model are motivated by UV completions of gravity such as string theory. We discuss the $\mathrm{SL}(2,\mathbb{R})$ and $\mathrm{O}(3)$ sigma models with interacting potentials and simulate their full nonlinear dynamics on black hole spacetimes. 
We study the properties of the field as a function of the curvature of the sigma model with respect to the free massive scalar case. In the accretion process, the  $\mathrm{SL}(2,\mathbb{R})$ model behaves as a self-interacting field with attractive interaction, while the $\mathrm{O}(3)$ one exhibits a repulsive phenomenology. In the case of a binary black hole system, these models cause a positive or negative dephasing of the gravitational waveform, respectively, when compared to the non-interacting case and as long as the field's Compton wavelength is larger than the binary separation. We observe no \textit{bosenova} emission, which suggests that nonlinearities tend to suppress this phenomenon. Our results highlight how the kinetic and potential terms are both relevant in determining the field dynamics.

\end{abstract}
\maketitle

\section{Introduction}

Gravitational waves provide a unique channel for probing gravity in a strong and dynamical regime. Besides uncovering a rich black hole (BH) population and allowing performing high-precision tests of General Relativity (GR) \cite{Maggio:2022hre,Pompili:2025cdc,Krishnendu:2021fga, LIGOScientific:2021sio,Carson:2019kkh,Cornish:2011ys}, these signals could carry information about the environments their sources are evolving in. Such environments can be of astrophysical nature, i.e. baryonic matter accretion disks, third-object interactions, or can be composed of dark matter (DM). In any of these cases, the environment can interact with binary compact objects, exchanging energy and angular momentum, thus modifying the inspiral behaviour with respect to a vacuum GR evolution \cite{Barausse:2014pra, Gliorio:2025cbh, Duque:2023seg, Ishibashi:2020zzy}. Achieving a detailed understanding of these effects is necessary, on the one hand, to characterise the environment and possibly extract new information about its constituents and, on the other hand, it is crucial to identify systematics and biases in the data analysis pipelines \cite{Cole:2022yzw,Zwick:2022dih}. 

In this work, we will focus on a DM environment composed by light scalar particles. Light scalar fields have been largely considered in the literature as possible DM candidates, as they emerge naturally from extensions to the Standard Model (SM) as well as from String Theory (ST) \cite{Arvanitaki:2009fg, Halverson:2017deq}.
The most common theory of this kind is the one of a complex scalar field $\Phi$  with a Lagrangian of the form 
\begin{equation}\label{eq:SM}
\mathcal{L}_{\Phi}=-\frac{1}{2}\partial_{\mu}\Phi\partial^{\mu}\bar{\Phi}-V\left(|\Phi|^2\right) .
\end{equation}

Such systems have been extensively discussed in the literature, usually taking the potential to be a mass term $V\sim\mu^2|\Phi|^2$. Such candidates have been shown to have a rich phenomenology when interacting with compact objects. For instance, they can build superradiant states around rapidly spinning BHs, which can in turn imprint specific signatures on the inspiral of a light secondary object \cite{Brito:2015oca,Baumann:2019eav,Baumann:2021fkf}. Recent works have shown that scalar fields can grow non-trivial over-densities around equal-mass binaries that are long-lived and induce a dephasing of the inspiral \cite{Bamber:2022pbs,Aurrekoetxea:2024cqd}. Different choices for the potential term have also been explored. Self-interactions of the scalar field have been shown to impact the total GW dephasing and, more generally, the system's phenomenology, with attractive self-interactions leading to a disruption of the cloud known as \textit{bosenova} \cite{Aurrekoetxea:2024cqd}. 

We will consider an extension to the theories \eqref{eq:SM}, which is given by promoting the kinetic term to a nonlinear sigma model,
\begin{equation}
\mathcal{L}_{\phi}=-\frac{1}{2}G_{AB}(\phi)\grad_{\mu}\phi^{A}\grad^{\mu}\phi^{B}-V\left(\phi^{A}\right).
\label{eq:general_sigma_model}
\end{equation}
In these theories, a set of scalars $\{\phi^A\}_{A=1,...,N}$ can be thought of as coordinates on a space called scalar manifold, with metric $G_{AB}$. In general, this space can be curved. Such models emerge naturally in supergravity and compactifications of string theories \cite{Burgess:1994kq, Brax:2023now, Cano:2023bpe}. This paper presents the implementation of such theories in the numerical-relativity (NR) code \texttt{GRChombo} \cite{Andrade:2021rbd, Radia:2021smk} and has the goal of characterising the influence of the sigma model parameters on the field's phenomenology in its full nonlinear regime. This takes advantage of the flexibility of this code in studying exotic physical systems in GR \cite{Figueras:2015hkb,Helfer:2018vtq,Figueras:2022zkg,Evstafyeva:2022bpr, Ge:2024itl,Marks:2025jpt} but also for the investigation of beyond-GR effects motivated by string theory or EFTs in the regime where nonlinear terms can be important \cite{Figueras:2020dzx,Figueras:2021abd,AresteSalo:2022hua,AresteSalo:2023mmd,Doneva:2023oww,Doneva:2024ntw,AresteSalo:2025sxc}.

The paper is organised as follows. In Section~\ref{sec:nonlinear} we introduce the nonlinear sigma models that will be the object of our work. In Section~\ref{sec:NR} we outline the implementation of the theory \eqref{eq:general_sigma_model} in the \texttt{GRChombo} code and we present the numerical results. In section~\ref{sec:results-accretion} we describe the evolution of such fields on the background of an isolated black hole and characterise the accretion process for various curvatures of the sigma model. In section~\ref{sec:results-binary} we treat the case of a binary black hole evolution in the presence of this type of scalar field. Finally, we conclude in \ref{sec:discussion} by discussing our results and outlining possible future directions.

We follow the conventions in Wald's book \cite{Wald:1984rg}. Greek letters $\mu,\nu\,\ldots$ denote spacetime indices and they run from 0 to $d$; Latin letters $i,j,\ldots$ denote indices on the spatial hypersurfaces and they run from 1 to $d$. We set $G=c=1$.

\section{Nonlinear sigma models }\label{sec:nonlinear}
In this section we introduce the two-dimensional nonlinear sigma models that are implemented and studied in the rest of this work. These models were derived in \cite{Cano:2023bpe} to study spherically symmetric soliton solutions (boson stars). 

One can start considering the axion-dilaton system which arises in supergravity and string theory \cite{Ortin:2015hya}. This model has the kinetic term 
\begin{equation}\label{eq:axidilaton1}
\mathcal{L}_{K}=-\frac{1}{2}(\partial\phi)^2-\frac{1}{2}e^{2\gamma\phi}(\partial a)^2\, ,
\end{equation}
where $\phi$ is the dilaton field, $a$ is the axion and $\gamma$ is a parameter determining the coupling between these scalars \footnote{Let us remark this theory is distinct from the QCD axion of \cite{Guerra:2019srj} as well as from axion-dilaton models from Gauss-Bonnet and dynamical Chern-Symons gravity recently investigated numerically in \cite{Richards:2025ows}.}. The kinetic term is invariant under $\mathrm{SL}(2,\mathbb{R})$ symmetry and the scalar manifold is thus the hyperbolic space. The symmetry can be made manifest after rewriting the Lagrangian in terms of the complex scalar field 
\begin{equation}\label{taudef}
\tau=\gamma a+i e^{-\gamma\phi}\, ,
\end{equation}
on which the $\mathrm{SL}(2,\mathbb{R})$  group acts via the M\"obius transformations
\begin{equation}
\tau'=\frac{c_1 \tau+c_2}{c_3 \tau+c_4}\, ,\qquad  \text{where} \, \,\,c_1c_4-c_2c_3=1\, ,
\end{equation}
with $c_i \in \mathbb{R}$. The kinetic terms \eqref{eq:axidilaton1} can be complemented with a mass term that breaks the symmetry to the $\mathrm{U}(1)$ subgroup, given by the transformation
\begin{equation}\label{eq:U1sym}
\tau'=\frac{\tau \cos(\alpha/2)+\sin(\alpha/2)}{\cos(\alpha/2)-\tau \sin(\alpha/2)}\, ,\quad  \alpha\in [0,2\pi)\, .
\end{equation}

This leads to a general theory given by
\begin{equation}\label{eq:axidilaton}
\mathcal{L}=-\frac{\partial_{\mu}\tau\partial^{\mu}\bar\tau}{2\gamma^2\text{Im}(\tau)^2}-U\left(\mathcal{T}^2\right),
\end{equation}
with 
\begin{equation}
\mathcal{T}^2=\frac{1}{\gamma^2}\left(\frac{1+|\tau|^2}{\text{Im}(\tau)}-2\right).
\end{equation}

The scalar manifold of the model \eqref{eq:axidilaton} is maximally symmetric, thus this is a maximally-symmetric sigma model. One can also construct the positive-curvature analogue with a sphere as the scalar manifold, subject to $\mathrm{O}(3)$ symmetry.  We can write the two theories in a unified form as follows. By changing variables into \req{eq:axidilaton} to
\begin{equation}
\Phi=\frac{2}{\gamma}\frac{1+i\tau}{1-i\tau}\, ,
\end{equation} 
the sigma model Lagrangian becomes
\begin{equation}\label{eq:axidilatonPhi}
\mathcal{L}=-\frac{\partial_{\mu}\Phi\partial^{\mu}\bar{\Phi}}{2\left(1-\frac{\gamma^2}{4}|\Phi|^2\right)^2}-\frac{\mu^2|\Phi|^2}{2\left(1-\frac{\gamma^2}{4}|\Phi|^2\right)}\, ,
\end{equation}
which is only dependent on $\gamma^2$. Then, with $\gamma^2>0$ the scalar manifold is the hyperbolic space, with $\mathrm{SL}(2,\mathbb{R})$ symmetry group, whereas with $\gamma^2<0$ it is the 2-sphere, $\mathrm{O}(3)$ symmetry. For $\gamma=0$ it reduces to the complex scalar field \req{eq:SM}.

The field $\Phi$ is complex and invariant under the $\mathrm{U}(1)$ symmetry, whose Noether charge is associated to particle number conservation. Complex scalar fields behaving asymptotically as $\Phi\sim\phi_0 e^{i\mu t}$ give rise to a stationary stress-energy tensor and, thus, stationary energy density. Scalar fields with these properties have been extensively studied, as they evade the no-hair theorem assumptions and allow the growth of a non-trivial profile around isolated BHs \cite{Jacobson:1999vr, Herdeiro:2015waa}. 

\section{Numerics}\label{sec:NR}

In this section, we discuss the implementation of the equations of motion arising from \eqref{eq:axidilatonPhi} in the \texttt{GRChombo} code. When varying the sigma model Lagrangian with respect to the field, one gets the Einstein-Klein Gordon equation 

\begin{align}
    \label{eq:EKG}
    &\grad_\mu\grad^\mu\Phi\nonumber\\&+\frac{\gamma^2\bar{\Phi}\grad_\mu\Phi\grad^\mu\Phi}{2\left(1-\frac{\gamma^2}{4}|\Phi|^2\right)}-2\Phi\left(1-\frac{\gamma^2}{4}|\Phi|^2\right)^2U'(|\Phi|^2)=0\,,
\end{align}
where we redefine the potential as dependent on $|\Phi|^2$ only for compactness and $U'(x)=\drm U/\drm x$. Varying the Lagrangian with respect to the metric yields the Einstein field equations with the stress-energy tensor associated to the field $\Phi$. In order to evolve this set of equations on a computer, we project them on a $(3+1)$ slicing following the Arnowitt-Deser-Misner formalism \cite{Arnowitt:1962hi}. We write the line element in the form 

\begin{equation}
    \drm s^2=-\alpha^2\drm t^2+\gamma_{ij}(\drm x^i+\beta^i\drm t)(\drm x^j+\beta^j\drm t)\,, 
\end{equation}
with $(\alpha,\beta^i)$ the lapse functions and shift vector and $\gamma_{ij}$ the $3$-dimensional spatial metric.\footnote{ Let us note here that the 3d metric $\gamma_{ij}$ should not be confused with the field's parameter $\gamma^2$, which will always appear as a squared scalar.} 
The field's stress energy tensor is decomposed in the components 

\begin{equation}\label{eq:SET-components}
\begin{aligned}
    &\rho=n_\mu n_\nu T^{\mu\nu}, &  &S_i=-\gamma_{i\mu}n_\nu T^{\mu\nu},\\
      &S_{ij}=\gamma_{i\mu}\gamma_{j\nu} T^{\mu\nu}, &  &S=\gamma^{ij}S_{ij}\,,
\end{aligned}
\end{equation}
with $n^\mu=(\partial_t^\mu-\beta^k\partial_k^\mu)/\alpha$ the future-directed unit-norm four-vector. We explicitly provide the equations for the stress-energy tensor quantities and the equations of motion of the field in the $(3+1)$ decomposition in Appendix \ref{sec:3+1}.

We obtain initial conditions that satisfy the Hamiltonian and Momentum constraints with the code \texttt{GRTresna} \cite{Aurrekoetxea:2025kmm}, which uses the recently proposed CTTK method \cite{Aurrekoetxea:2022mpw} based on the standard Conformal Transverse-Traceless decomposition. In particular, we use the ``hybrid CTTK'' method (see \cite{Aurrekoetxea:2022mpw} for further details) to solve the constraints equations, which enables us to obtain second-order convergent initial data (see Appendix \ref{sec:convergence}).

For the runs with single BHs we use a computational domain of $L = 512M$ with the BH situated at the centre of the grid, and $N=64$ grid points on the coarsest level. We employ $9$ levels of refinement with a refinement ratio of $2:1$, which results in a finest resolution of $dx_{finest} = M/32$ on the finest grid, giving $\sim 75$ grid points across the BH horizon in the quasi-isotropic Kerr coordinates \cite{Liu:2009al} that we use to set the initial conditions for the metric.

For the BBH mergers we have chosen $L=1024 M$, with $N=128$ grid points on the coarsest level, with $10$ levels of refinement, which results in a resolution of $dx_{finest} = M/64$ on the finest grid, which gives again roughly $75$ points across the horizon of each BH prior to their merger. 

For both type of simulations we use $4^\text{th}$ order finite differences to discretise the spatial derivatives and a standard $4^\text{th}$ order Runge-Kutta time integrator to step forward in time.

\subsection{Isolated black hole}\label{sec:results-accretion}

\begin{figure*}[t!]
	\centering
	\includegraphics[width=.95
    \textwidth]{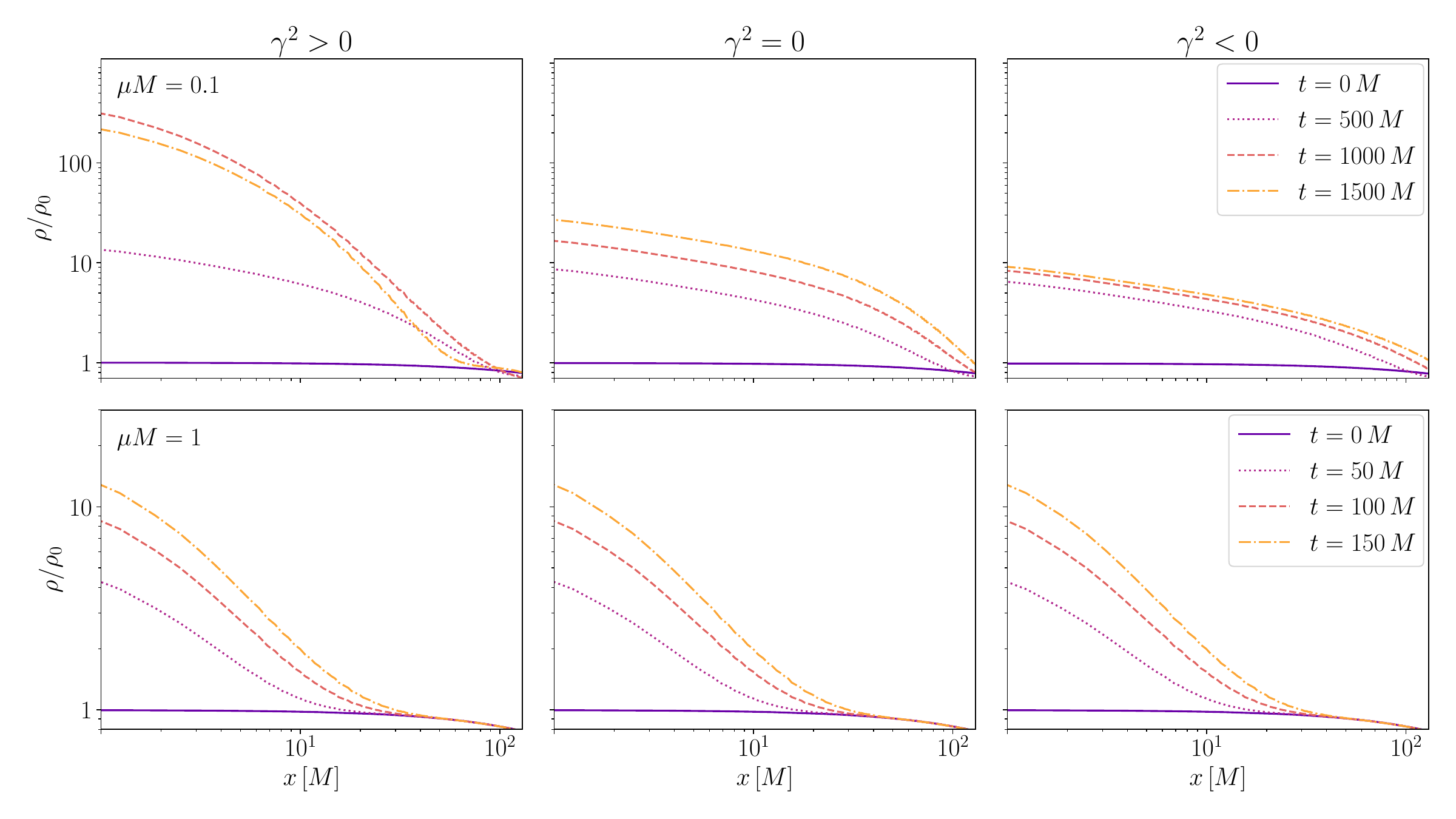}
	\caption{Radial density profile around a Schwarzschild BH for different values of $\gamma^2$. The top panels are obtained with a scalar field mass of $\mu M=0.1$. The bottom panels correspond to a scalar mass of $\mu M=1$.}
	\label{fig:accretion_SS}			
\end{figure*}

We simulate the evolution of the scalar field \eqref{eq:axidilatonPhi} around a Schwarzschild BH for different masses of the scalar field as well as sigma model curvatures. We take as initial condition a uniform field distribution, such that $\Phi|_{t=0}=\phi_0$, with $\phi_0\in \mathbb{R}$, and $\Pi|_{t=0}=-i\mu \phi_0$. When varying the scalar field mass $\mu$, we adjust the field amplitude so that the initial density stays of order $\rho|_{t=0}\sim\mu^2\phi_0^2 \sim 10^{-9} M^{-2}$. This value is consistent with previous numerical work on scalar fields on BH spacetimes \cite{Clough:2019jpm, Aurrekoetxea:2024cqd} and it is comparable to matter densities in DM halos grown adiabatically or through superradiance, while larger than current constraints of DM densities in galactic environments. We choose extrapolating boundary conditions with a first-order scheme for the scalar field variables, to allow a non-zero oscillating value at infinity. This allows for a superposition of ingoing and outgoing modes. We checked that this choice of boundary conditions did not significantly impact the simulations results for different sizes of the simulation box, and we used a large domain length to minimise boundary artifacts. We plot the radial density profile in the vicinity of the BH at different times in Fig. \ref{fig:accretion_SS}. We choose the values of $\gamma^2$ to be $\pm 10^5$ so that $M^2|\gamma^2|\rho|_{t=0}\sim 3$, which mimics the same behaviour as in \cite{Cano:2023bpe}.

In all cases, the scalar field accretes on the central BH at early times and the density grows in the inner region of the domain. For a small field mass $\mu M=0.1$, the $\mathrm{SL}(2,\mathbb{R})$ sigma model, with $\gamma^2>0$, produces a denser and more compact distribution than the trivial case. On the other hand, the $\mathrm{O}(3)$ model, $\gamma^2<0$, yields lower central densities and a spread-out profile. Thus, $\gamma^2>0$ gives an attractive self-interacting dynamics that enhances the field accretion, whereas $\gamma^2<0$ shows a repulsive self-interaction. For a heavier mass of the scalar field $\mu M=1$, the three cases generate the same accretion profile. The mass term dominates the dynamics over the nonlinear self-interactions induced by the scalar manifold's curvature. For the heavier mass simulations, we use a time step given by $dt_\text{multiplier}=0.0625$ (with $\Delta t_{coarsest}=L/N\cdot dt_\text{multiplier})$,\footnote{This is four times smaller than the one we generally use ($dt_{multiplier}=0.25$), which is given by the Courant–Friedrichs–Lewy (CFL) condition of our discretisation scheme.} which allows to resolve the scalar field oscillations on the coarsest level.

\subsection{Binary black hole}\label{sec:results-binary}

We simulate a binary black hole (BBH) with BHs of equal masses $m_{(1)} = m_{(2)} = 0.48847892320123M$, initial separation $12.21358M$ and initial momenta $p_{(i)}/M = (\pm 0.0841746, \mp 0.000510846, 0)$. These initial conditions were tuned to have quasi-circular initial orbits in the absence of a dark matter cloud, a total ADM mass of $M_{ADM}=M$, and such that the two black holes merge in approximately ten orbits. We superpose the solutions for two boosted black holes as described in \cite{Baumgarte:2010ndz,Bowen:1980yu}, using a perturbative solution for the conformal factor that is accurate up to order $(P^i P_i)^2$, and we choose again an initial uniform distribution for the scalar field such that $(\Phi_{t=0},\Pi_{t=0})=(\phi_0,\,-i\mu\,\phi_0)$. Then we solve the Hamiltonian and momentum constraints using the CTTK approach as mentioned above.

When studying the dephasing of the waveforms in the different BBH systems considered, we have chosen to align them at a given frequency in the same way as in \cite{AresteSalo:2025sxc}, which enables us to get more consistent results which are less affected by the ejection of the junk radiation. The frequency of the gravitational wave as a function of time can be estimated by computing
the gradient of the trajectories of the punctures or the gradient of the phase of the gravitational
wave and we have chosen the initial frequency to be $f_0=0.00716/M$, which corresponds to the time slice that we refer to as $t_0$.

\subsubsection{$\mu M=0.3$}
\label{sec:mu03}

\begin{figure}[htbp]
	\centering
	\includegraphics[width=0.5\textwidth]{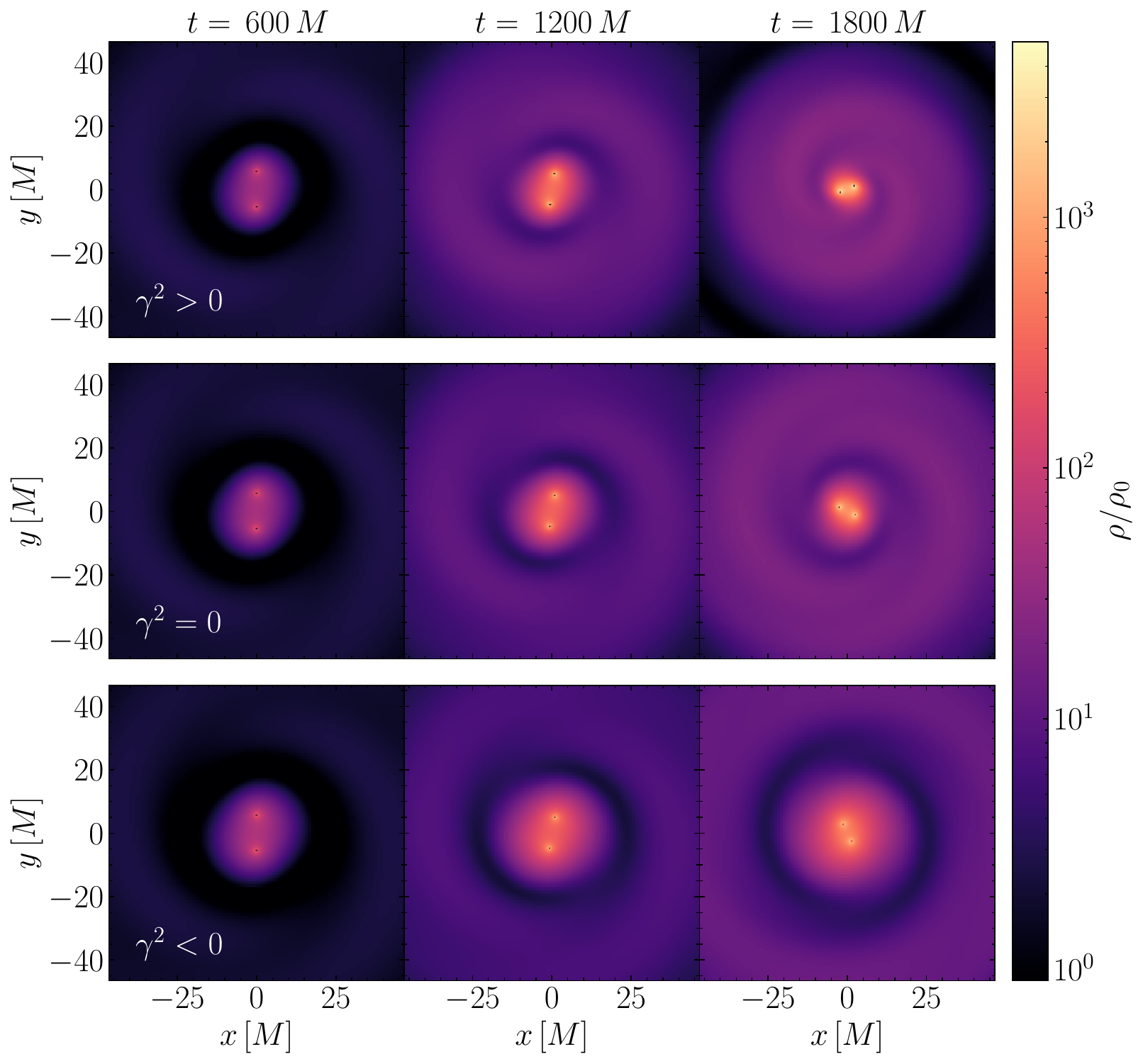}
	\caption{Scalar field density evolution around an equal-mass BBH for $\mu M=0.3$ and $\gamma^2=10^5$ (top), $\gamma^2=0$ (middle), $\gamma^2=-10^5$ (bottom).}
	\label{fig:binary_mu03_density}			
\end{figure}

We first consider a scalar field mass of $\mu M=0.3$. The Compton wavelength of the field $\lambda_C=2\pi/\mu \sim 20$ is then larger than the initial separation of the binary. We run three simulations with the same initial conditions and $\gamma^2\in\{10^5,0,-10^5\}$ respectively. We plot the density evolution on the $z=0$ plane in Fig. \ref{fig:binary_mu03_density}, normalised with respect to the uniform density at time $t=0$.

For all scalar manifold's curvatures, the field builds up an over-density around both BHs, that grows in magnitude as the binary shrinks. Comparing the distributions at late times, the case $\gamma^2>0$ forms a compact cloud around the BBH, that rapidly decays with the radius to lower density values. On the other hand, $\gamma^2<0$ gives rise to a more diffused cloud, which reaches lower densities in the region in between the binary components and decays more slowly with the radius. This would suggest that $\gamma^2>0$ behaves like an attractive self-interacting field, whereas $\gamma^2<0$ displays a repulsive self-interaction, resulting in a spread-out distribution. We remind that the $\gamma^2=0$ case reduces to a non-interacting, minimally-coupled massive scalar.

\begin{figure}[htbp]
	\centering
	\includegraphics[width=0.5\textwidth]{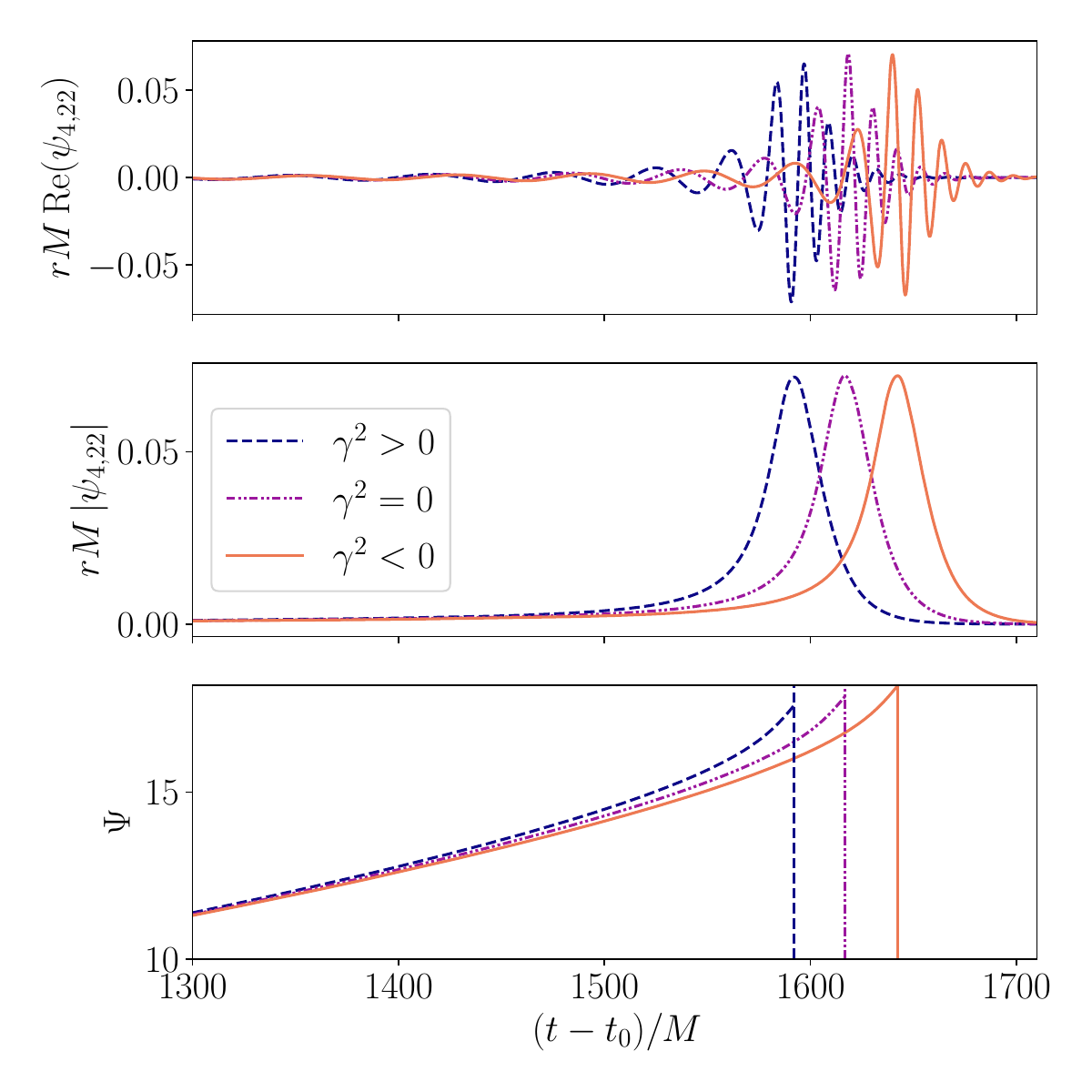}
	\caption{GW emission of an equal-mass BBH for $\mu M=0.3$. The top panel shows the real part of the Weyl scalar $\psi_4$ extracted at $r=50 M$, while the middle panel plots its absolute value. The bottom panel shows the accumulated orbital phase $\Psi$.}
	\label{fig:binary_mu03_combined}			
\end{figure}

The impact of the sigma model curvature on the inspiral is displayed in Fig. \ref{fig:binary_mu03_combined}. Both positive and negative curvatures induce a dephasing of the GW emission with respect to the non-interacting case. The positive curvature case merges first, in agreement with the intuition that $\gamma^2>0$ corresponds to attractive self-interactions. In the same way, $\gamma^2<0$ makes the inspiral phase last longer than in the non-interacting case. The bottom panel of the figure shows the accumulated orbital phase in the three inspirals, which we define as half the complex phase of the $l=m=2$ mode of the Weyl scalar $\psi_4$.

\subsubsection{$\mu M=0.6$}

\begin{figure}[htbp]
	\centering
	\includegraphics[width=0.5\textwidth]{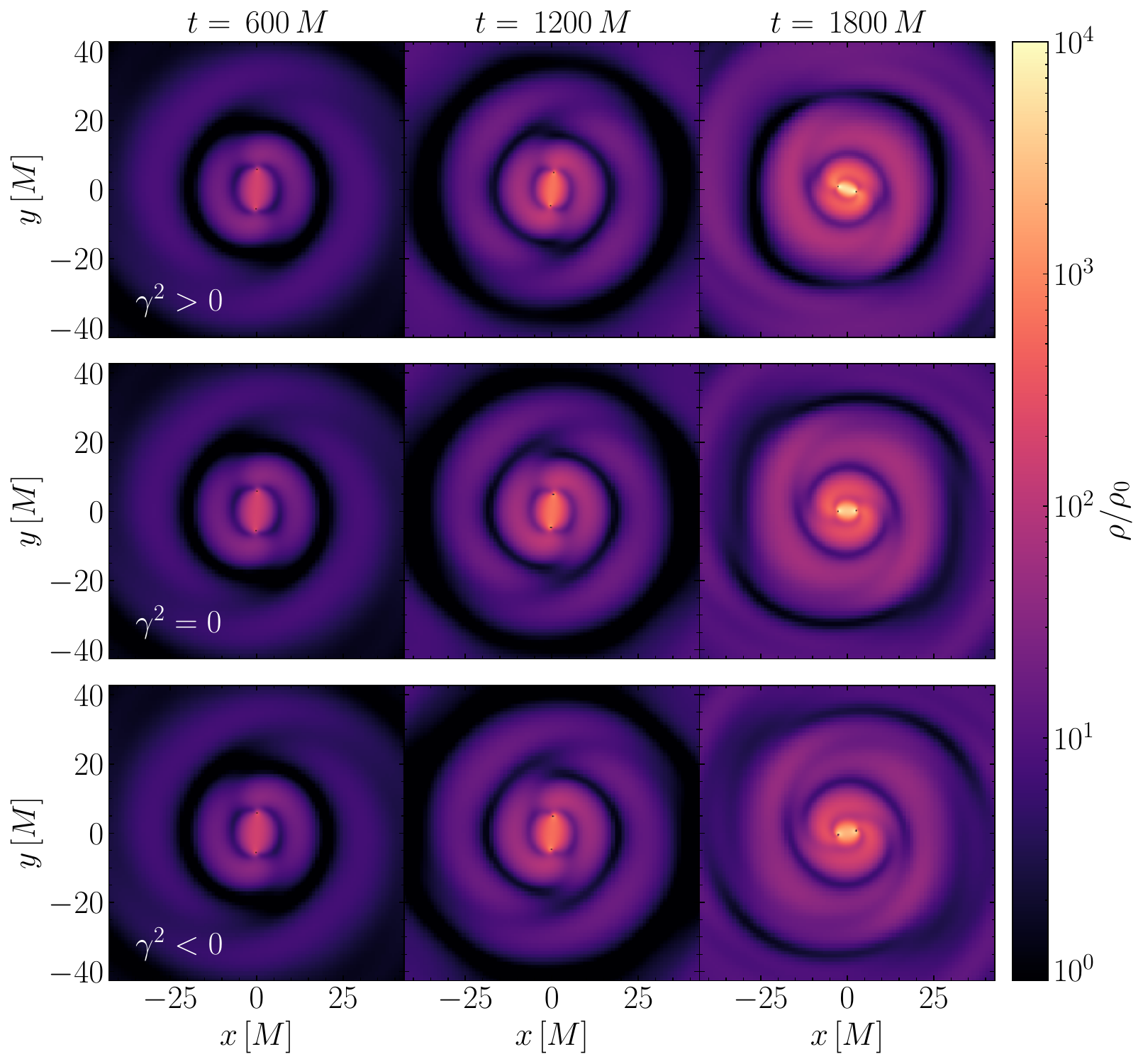}
	\caption{Scalar field density evolution around an equal-mass BBH for $\mu M=0.6$ and $\gamma^2=10^5$ (top), $\gamma^2=0$ (middle), $\gamma^2=-10^5$ (bottom).}
	\label{fig:binary_mu06_density}			
\end{figure}

We perform the same analysis as the one presented in section \ref{sec:mu03} with a scalar field of double the mass, i.e. we now consider $\mu M=0.6$. The Compton wavelength is thus $\lambda_C \sim 10M$, comparable to the initial separation of the BHs. In Fig. \ref{fig:binary_mu06_density} we plot the scalar field density at different times of the evolution for the same three choices of $\gamma^2$ as in the previous section. A first inspection of the density distributions shows a more complex cloud structure than for the light particle cases. The shorter Compton wavelength imprints a stronger wave-like behaviour to the field at scales commensurate to the BHs orbital radii.  

\begin{figure}[htbp]
	\centering
	\includegraphics[width=0.5\textwidth]{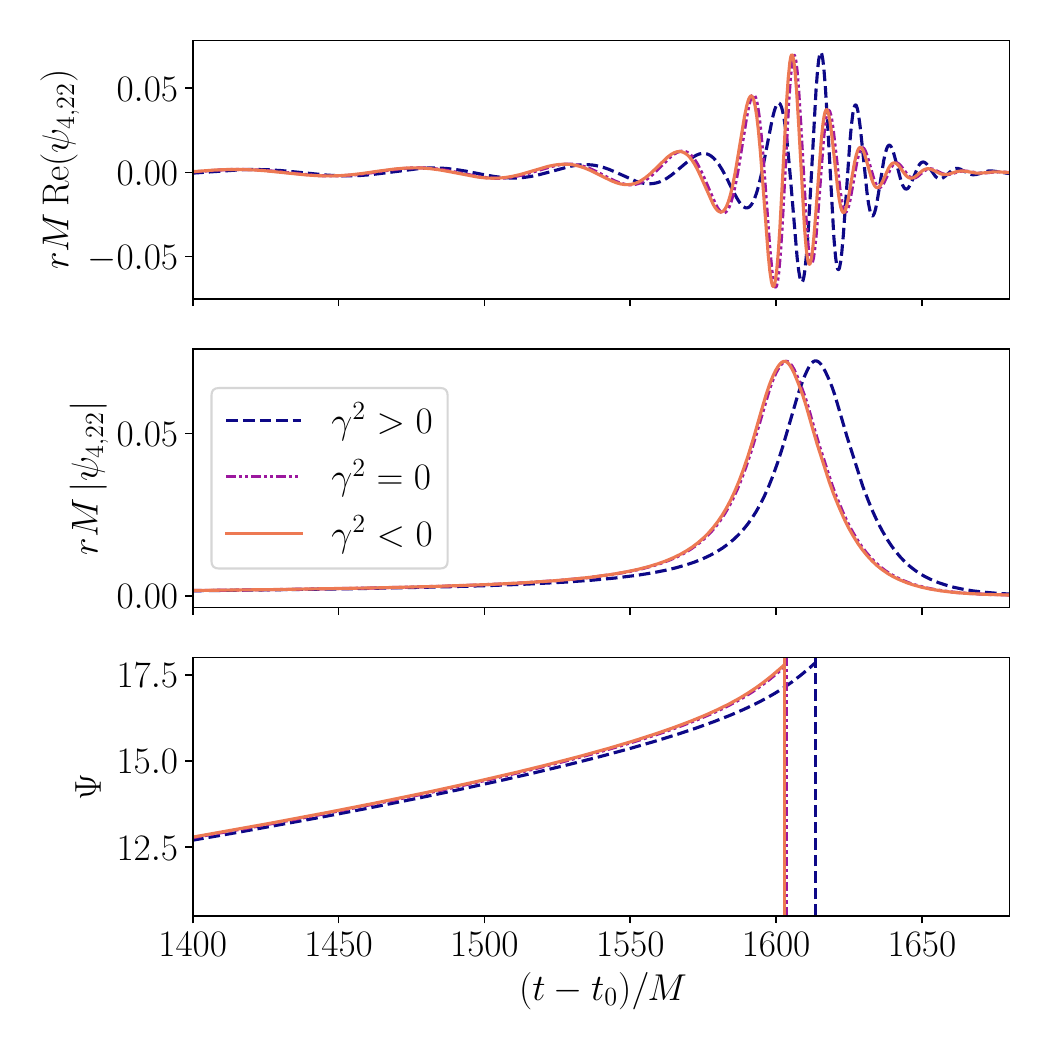}
	\caption{GW emission of an equal-mass BBH for $\mu M=0.6$. The top panel shows the real part of the (2,2) mode of the Weyl scalar $\psi_4$ extracted at $r=50 M$, while the middle panel plots its absolute value. The bottom panel shows the accumulated orbital phase $\Psi$. }
	\label{fig:binary_mu06_combined}			
\end{figure}

The field's mass impacts significantly the inspiral dynamics and the phenomenology of the emitted GWs. These are plotted in Fig. \ref{fig:binary_mu06_combined}. Contrary to the results of section \ref{sec:mu03}, the $\gamma^2>0$ field yields now the longest inspiral time, while $\gamma^2=0$ and $\gamma^2<0$ merge at the same time. 

\begin{figure}[htbp]
	\centering
	\includegraphics[width=0.5\textwidth]{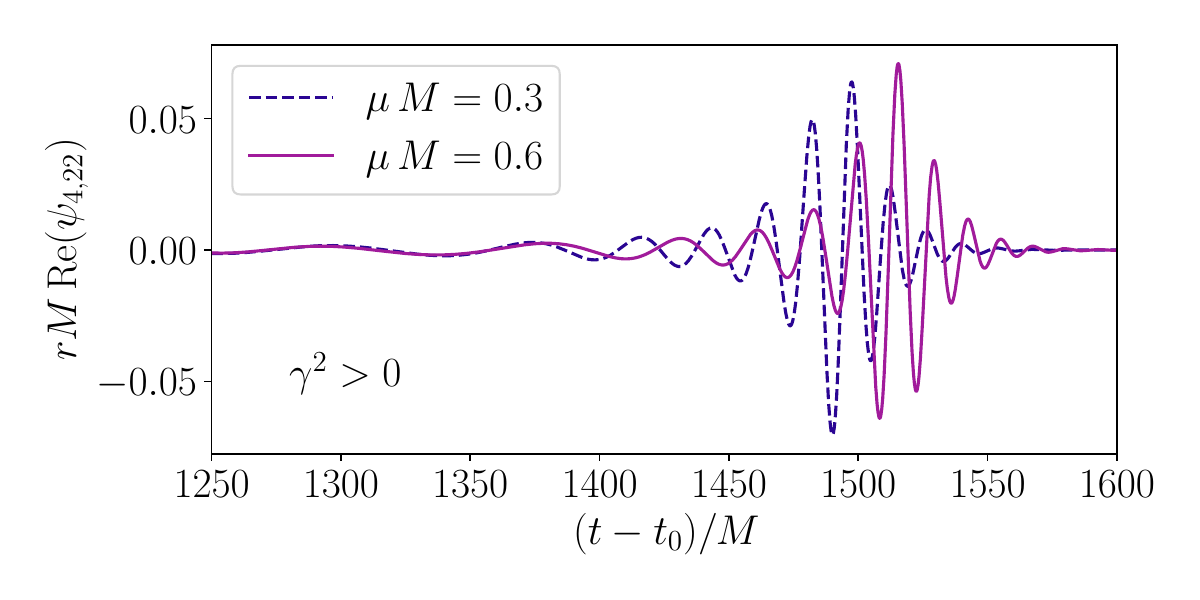}
	\caption{Real part of the (2,2) mode of the Weyl scalar for $\mu M=0.3$ (dashed line) and $\mu M=0.6$ (continuous line) with $\gamma^2=10^5$. The binary interacting with heavier field merges at a later time.}
	\label{fig:binary_2masses_gPos}			
\end{figure}

This result demonstrates the importance of self-interactions and non-standard kinetic term on the field dynamics and its impacts on a BBH evolution. Recent work \cite{aurrekoetxeaEffectWaveDark2024} showed that for a binary with a non-interacting massive scalar field, the dephasing is maximised when the Compton wavelength is comparable to the initial orbital separation $\lambda_C\sim d$. For the sigma model with positive curvature, this does not hold true: Fig. \ref{fig:binary_2masses_gPos} compares the GW emitted by the binary with the two choices of field's masses. The heavier field corresponds to a longer inspiral phase.

\begin{figure}[htbp]
	\centering
	\includegraphics[width=0.49\textwidth]{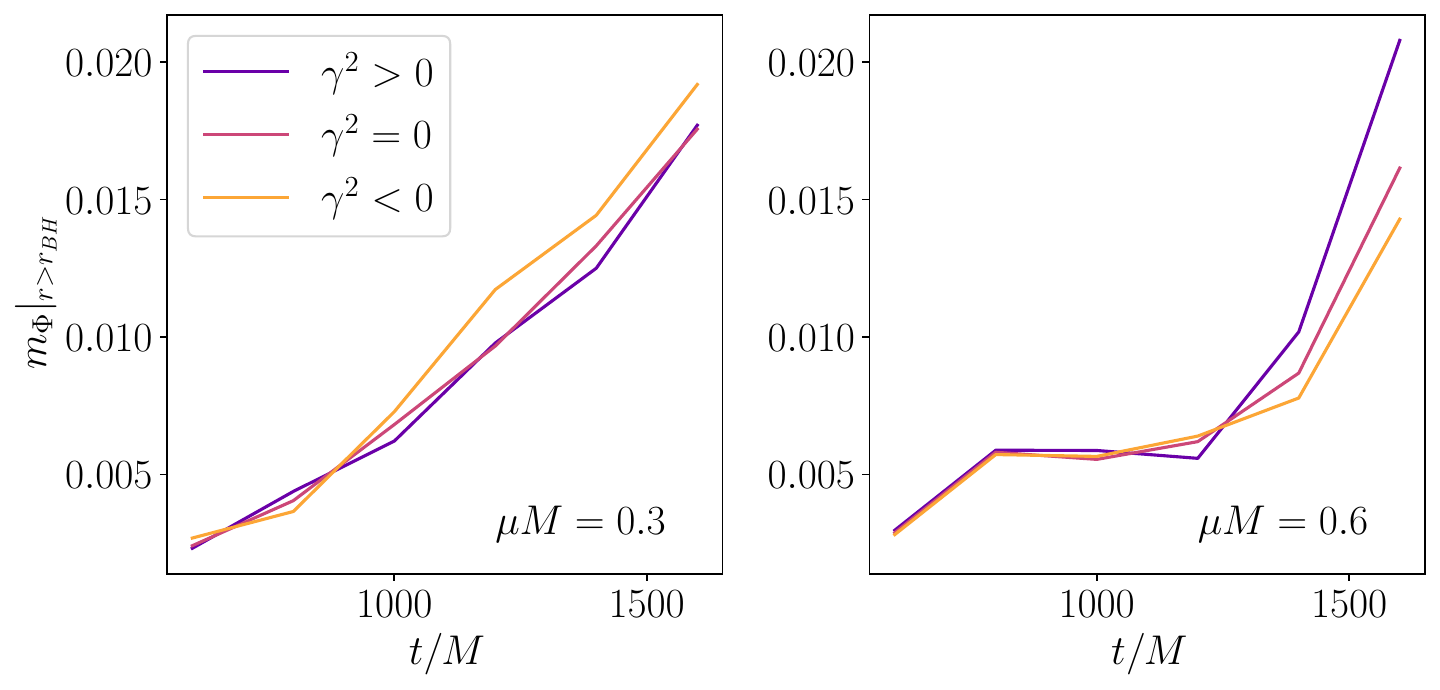}
	\caption{Integrated scalar field mass $m_{\Phi}$ in the region $r_{BHs}<r< 100 M$ plotted as a function of the inspiral time for $\mu M=0.3$ (left) and $\mu M=0.6$ (right), as well as for the chosen scalar manifold curvatures.}
	\label{fig:binary_mass_integral}			
\end{figure}

The complex structure and density distribution of the scalar cloud can therefore exert a radial pull on the binary components, competing with dynamical friction in driving the inspiral. We can test this idea by computing the scalar field total mass in the outer region of the system, i.e. for $r> r_{BHs}$ with $r_{BHs}$ the distance of the binary components from the centre of mass, namely $r_{BHs}=\sqrt{r_1^2+r_2^2}$, where $r_i$ is the distance of every black hole from the centre of mass. We estimate the mass via the volume integral 

\begin{equation}
    m_{\Phi}=-\int_\mathcal{V}\rm{d}^3x\,\sqrt{\bar\gamma}\,T_0^0\sim\int_\mathcal{V}\rm{d}^3x\,\sqrt{\bar\gamma}\,\alpha\,\rho\,,
\label{eq:mass_integral}
\end{equation}
in an analogous way as in \cite{Aurrekoetxea:2024cqd}, where $\bar\gamma$ is the determinant of the spatial metric. In Fig. \ref{fig:binary_mass_integral} we plot the resulting mass as a function of the inspiral time for the considered masses and sigma models. In both lighter and heavier cases, the curve with higher mass in the late part of the inspiral merges last. The additional mass distributed around the binary because of the non-trivial kinetic and potential terms is therefore an important driver of the inspiral physics, together with dynamical friction. 

In the BBH simulations we use a time step given by $dt_\text{multiplier}=0.25$, which is likely to introduce some errors in resolving the scalar field oscillations at the coarsest level. This value was used to keep the simulations running in the available computing time. We estimate that this limitation does not hinder significantly the results, as the code converges properly as discussed in appendix \ref{sec:convergence}.

\subsubsection{Ringdown phase}

\begin{figure}[htbp]
	\centering
	\includegraphics[width=0.5\textwidth]{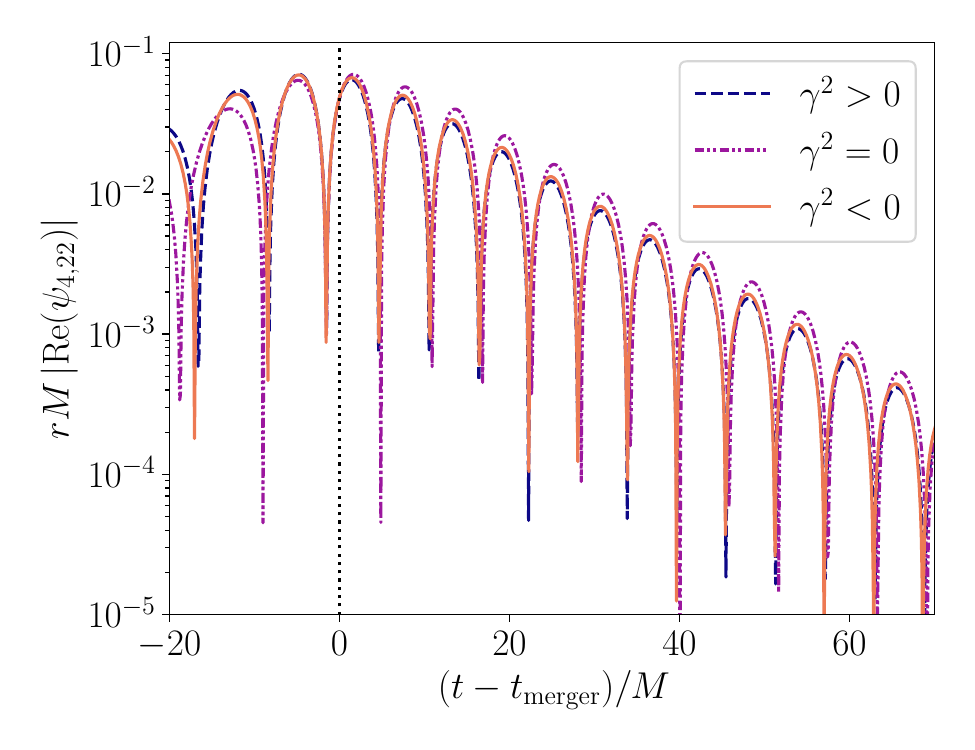}
	\caption{Merger and ringdown phases for a BBH with scalar field of $\mu M=0.3$ and different $\gamma^2$. We align the GWs at the maximal amplitude. The Weyl scalar is extracted at $r_{ex}=50 M$.}
	\label{fig:binary_mu03_QNMs}		
\end{figure}

The frequency content of the ringdown stage is of key importance to perform high-precision tests of the Kerr hypothesis \cite{Ferrari:2007dd, Dreyer:2003bv} and can receive corrections from beyond-GR extensions \cite{Pierini:2021jxd,Cano:2023jbk,Pezzella:2024tkf, Chakraborty:2024gcr}. In Fig. \ref{fig:binary_mu03_QNMs} we plot the absolute value of $\psi_{4,22}$ with $\mu M=0.3$ and varying $\gamma^2$. In order to study the impact of the sigma models on the quasi-normal modes (QNMs) of the simulated binaries, we align the emitted waveforms at merger and fit an exponentially damped oscillating function to the $l=2,m=2$ component of $\psi_4$, 

\begin{equation}
    \psi_{4,22}=C+A\,e^{- w_1\,t}\cos(w_2\,t+B).
    \label{eq:QNM_fit}
\end{equation}

We perform the fit starting from $t-t_{merger}\sim30M$, so to avoid the nonlinear contributions from the merger, up to $t-t_{merger}\sim 70M$. For all choices of $\mu M$ and $\gamma^2$, we obtain frequencies consistent with 

\begin{equation}
    (\omega_R=0.542\pm0.002,\,\omega_I=-0.083\pm0.001)\,,
    \label{eq:QNM}
\end{equation}
with the uncertainties coming from fits on simulations with different resolutions, as discussed in Appendix \ref{sec:convergence}. We can compare those values with the analytical (2,2) mode of a Kerr vacuum BH with dimensionless spin $\chi=0.69$, which is the final state of our simulations. The fitted QNMs differ from the reference value $\omega^{Kerr}_{22}=0.527-0.081\,i$ \cite{Berti:2009kk} of $\sim2\%$. The parameters explored with our simulations are thus insufficient to draw any conclusions about possible corrections to the QNM frequencies due to the sigma model curvature.

\begin{figure}[htbp]
	\centering
	\includegraphics[width=0.5\textwidth]{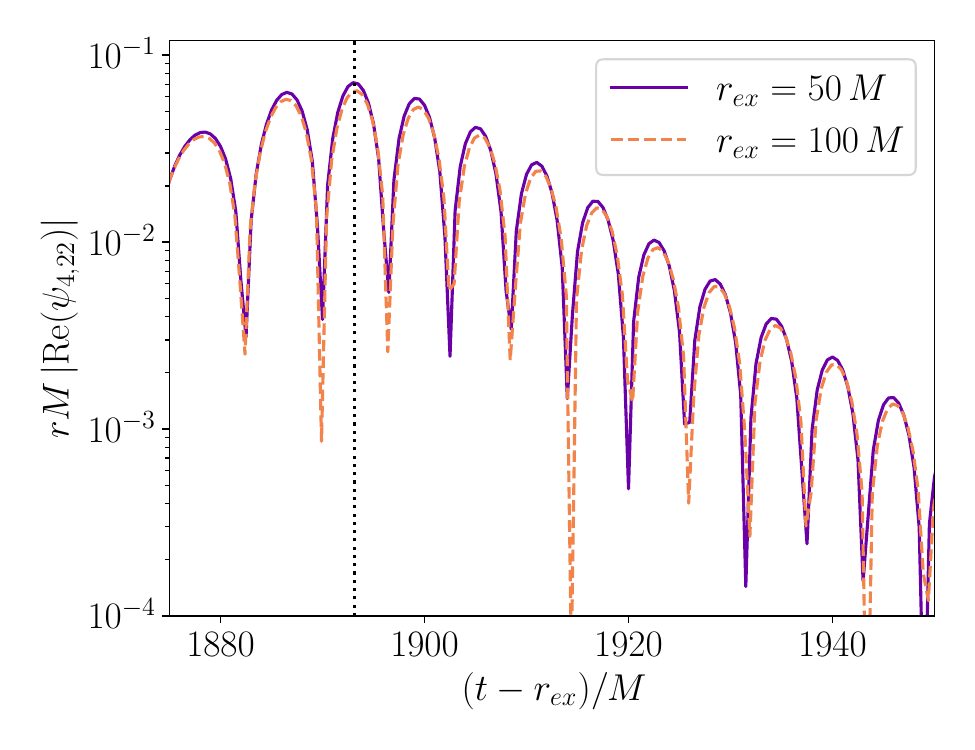}
	\caption{Waveform of a BBH with scalar field of $\mu M=0.3$ and $\gamma^2=0$ extracted at different radii as a function of $t-r_{ex}$. The waveforms align, showing no velocity dispersion in the gravitational radiation.}
	\label{fig:binary_QNMs_extr}
\end{figure}

In Fig. \ref{fig:binary_QNMs_extr} we plot the norm of the real part of the (2,2) mode of the Weyl scalar for $\mu M=0.3$ and $\gamma^2=0$ at two different extraction radii, $r_{ext}=(50M, 100M)$. We plot the GW against the retarded time after aligning the waveforms at some early-inspiral phase. The curves superimpose, showing that the tensor modes propagate with speed $1$ on the simulation grid.

\subsubsection{Scalar emission}

\begin{figure}[htbp]
\centering
\subfigure[]{\label{fig:binary_scalar_a}\includegraphics[width=.5\textwidth]{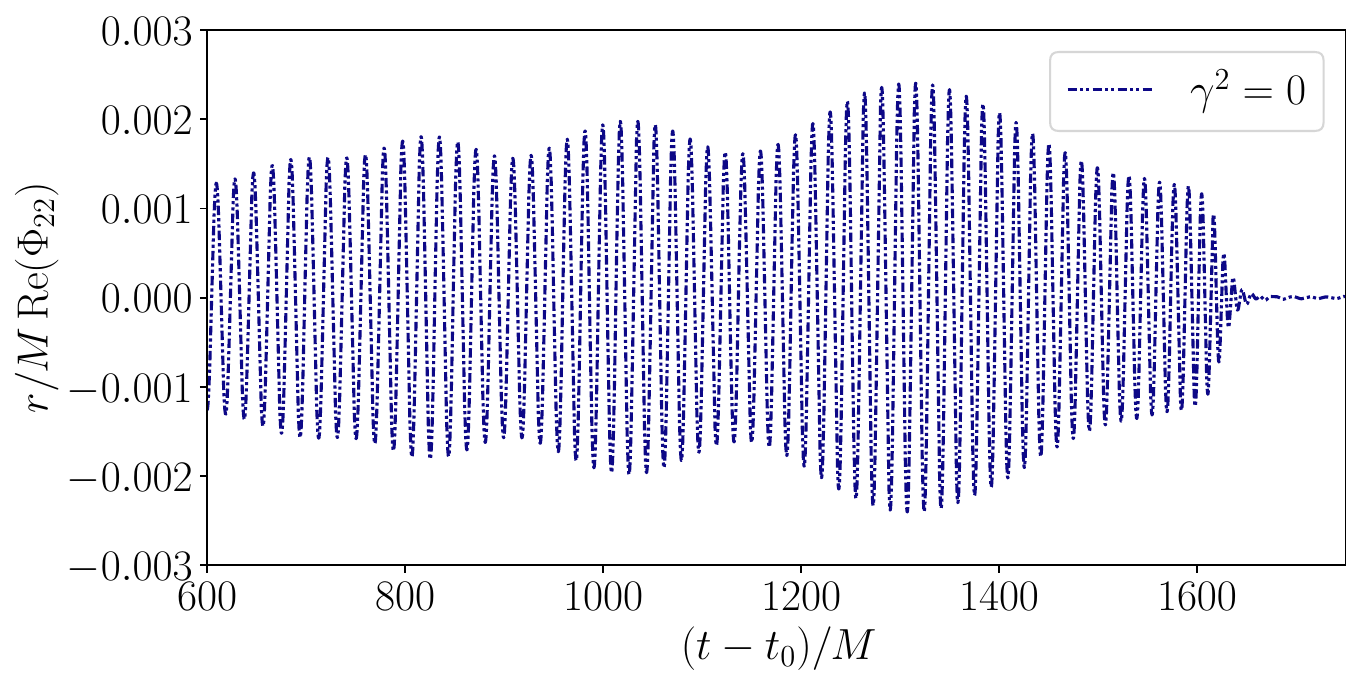}}
\medskip
\subfigure[]{\label{fig:binary_scalar_b}\includegraphics[width=.5\textwidth]{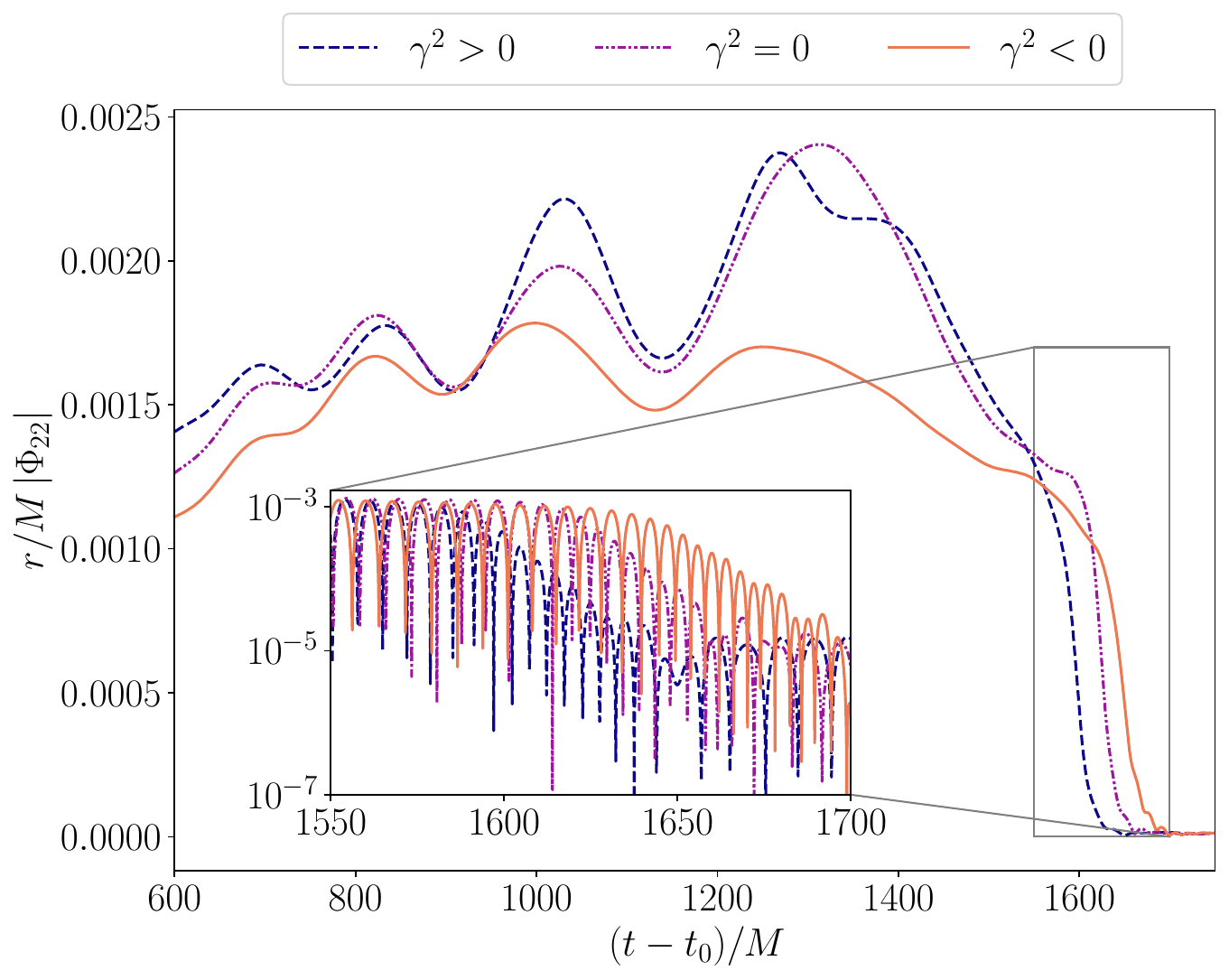}}
\caption{
(a) Scalar radiation from the $l=2,\,m=2$ mode with $\mu M=0.3$ and $\gamma^2=0$. The field shows a time-dependent maximal amplitude. 
(b) Absolute value of $\Phi_{22}$ for different values of $\gamma^2$. The inset plots the absolute value of $\mathrm{Re(\Phi_{22})}$ around the time of merger. }
\end{figure}

We characterise the scalar field dynamics extracting the scalar field value at several radii from the BBH centre of mass and projecting it on spherical harmonics. We focus our discussion on the behaviour of the $l=2,m=2$ mode, as it is the dominant one sourced by the quadrupolar GW emission. We also observe an oscillating $l=0,m=0$ mode, which contributes to the monopolar accretion on the binary and it is long-lived after the merger, as the field keeps accreting on the remnant. The massive nature of the field makes its propagation non-trivial \cite{Aurrekoetxea:2022ika}. In Fig. \ref{fig:binary_scalar_a} we plot the real part of the $(2,2)$ mode of the complex scalar field $\Phi$ extracted at $r_{ex}=50M$. The field oscillates rapidly during the inspiral. 

The field exhibits a varying maximal amplitude that grows at first, before tapering down and transitioning to a short exponentially damped ringdown. In Fig. \ref{fig:binary_scalar_b} we plot the field's absolute value for the different choices of $\gamma^2$. As already shown by the energy-density distributions, the scalar amplitude is strongly dependent on the field interactions. 

\begin{figure}[htbp]
	\centering
	\includegraphics[width=0.5\textwidth]{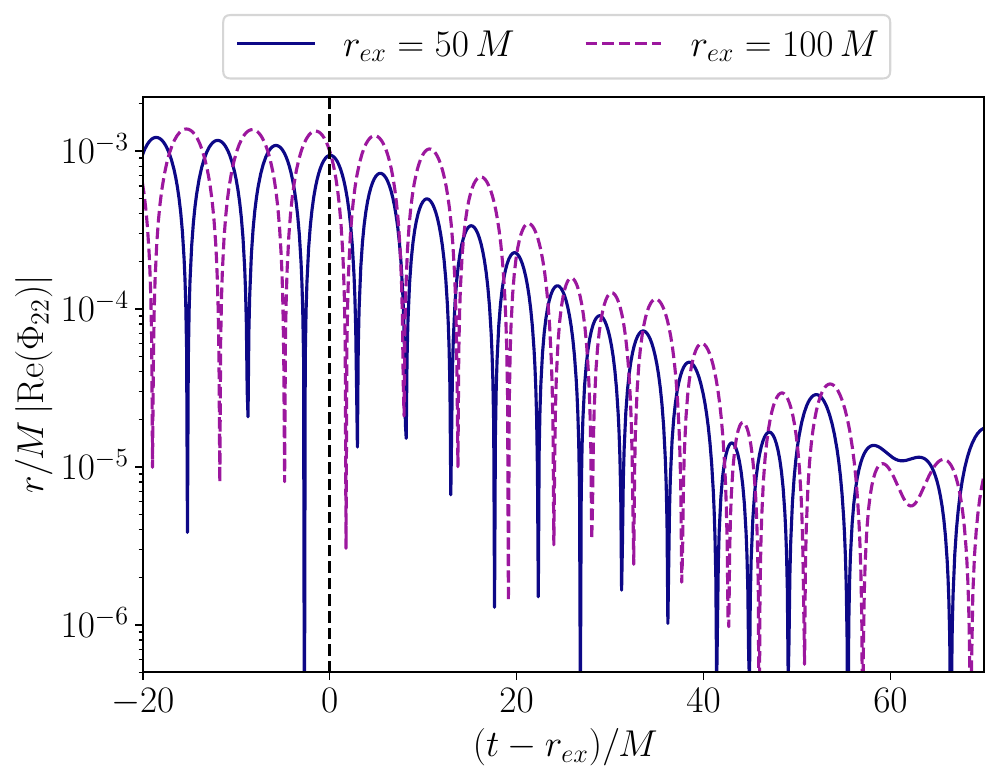}
	\caption{Scalar waveform of the $(2,2)$ mode at extraction radii $r_{ex}=50 M$ and $r_{ex}=100 M$. We plot the results for $\gamma^2=0$.}
	\label{fig:binary_scalar_ext}
\end{figure}

Moreover, the field mass introduces a translation into a frequency-dependent propagation velocity for the modes making up the emission. In Fig. \ref{fig:binary_scalar_ext} the scalar waves are extracted at different radii and plotted as a function of the retarded time. The curves do not match, as different frequencies components propagate at different speeds.

\section{Discussion}\label{sec:discussion}

In this work, we numerically studied the dynamics of the 2-dimensional maximally symmetric sigma models $\mathrm{SL}(2,\mathbb{R})$ and $\mathrm{O}(3)$. We evolved the theories' equations of motion in the code \texttt{GRChombo}, starting from constraints-obeying initial data obtained from \texttt{GRTresna}. 
We characterised the accretion of a uniform scalar field onto a non-rotating BH. For light values of the field mass, the $\mathrm{SL}(2,\mathbb{R})$ model exhibits attractive self-interaction with respect to a non-interacting massive field, with a larger and more compact over-density. The $\mathrm{O}(3)$ instead shows a repulsive behaviour. For larger values of the field mass, however, the mass term in the potential dominates over the self-interactions and the accretion is independent of the sigma model curvature. Our simulations show that scalar fields from sigma models can grow clouds from accretion over short timescales. The final state of such halos depends on the sigma model curvature as well as on the field mass. 

We evolved an equal-mass binary BH inspiral for $\sim10$ orbits, with different scalar manifold's curvatures and field mass, starting with the same initial separation and momenta for the BHs and background scalar field density. When the field Compton wavelength is larger than the initial BH separation, the field grows a circumbinary cloud during the inspiral. The cloud is more compact for the $\mathrm{SL}(2,\mathbb{R})$ model while it is more diffused in the $\mathrm{O}(3)$ case. The matter distribution effect on the binary results in a dephased GW emission. The $\mathrm{SL}(2,\mathbb{R})$ model extracts more energy from the BBH, which then merges earlier than in the non-interacting case. The $\mathrm{O}(3)$ model yields a longer inspiral than in the non-interacting case.

We therefore show the relevance of the non-trivial kinetic term for the field's phenomenology in astrophysical environments. Expanding the potential term at first order in $\gamma^2$ is equivalent to a $\lambda/4|\Phi|^4$ self-interaction potential with $\lambda = \mu^2\gamma^2/2$. However, contrary to the results given in \cite{Aurrekoetxea:2024cqd}, with $\lambda<0$ inducing attractive self-interactions, here it is the $\gamma^2>0$ case to exhibit such behaviour. We also point out that we do not observe a cloud disruption from \emph{bosenova} explosion, as reported in \cite{Aurrekoetxea:2024cqd} in the attractive case. This result suggests that the curvature dependence of the kinetic term suppresses the \emph{bosenova} onset. A direct comparison with the parameters and result of \cite{Aurrekoetxea:2024cqd} is spoiled by the dependence on $\gamma^2$ of our kinetic term. More simulations with strong attractive self-interaction are therefore needed to better understand such phenomenon.

Our results are consistent with the conclusions of \cite{Cano:2023bpe} characterising the soliton solutions of \eqref{eq:axidilatonPhi}. The $\mathrm{O}(3)$ would support heavy and compact boson stars, while the $\mathrm{SL}(2,\mathbb{R})$ case would lead to light and diffuse stars. Scalar fields with a standard kinetic term and repulsive self-interactions have been shown to produce boson stars with large maximal mass, of order $\sqrt{\lambda}M_P^3/\mu^2$, with $M_P$ the Planck mass, while attractive self-interactions would produce significantly lighter stars, with maximal masses of order $M_P/\sqrt{|\lambda|}$ \cite{Eby:2015hsq, Visinelli:2021uve}. In our numerical simulations, the $\mathrm{O}(3)$ ($\mathrm{SL}(2,\mathbb{R})$) model displays again a repulsive (attractive) self-interaction phenomenology. 

When thinking of the scalar field we simulate as a DM constituent, its physical mass $m=\hbar \mu/c$ is effectively unconstrained over a large domain by cosmological and astrophysical observations, with a lower bound at $m\sim10^{-21} \,\rm{eV}$ \cite{Hui:2021tkt}. The scalar's self-interactions, however, are constrained by dwarf galaxies and early universe dynamics \cite{Li:2013nal,Diez-Tejedor:2014naa}. The dimensionless self-interaction coupling is constrained as $\lambda\lesssim\left(m/\rm{eV}\right)^4$. For the mass used in our simulations, 
\begin{equation}
    \mu\sim0.5 M \Rightarrow\,m\sim8\cdot10^{-17}\left(\frac{M}{10^6M_\odot}\right)^{-1}\,\rm{eV},
\end{equation}
this translates to a bound of 
\begin{equation}
    \lambda\lesssim10^{-65}\left(\frac{M}{10^6M_\odot}\right)^{-4}.
\end{equation}
Then, for $|\gamma^2|=10^5$ we get
\begin{equation}
    \lambda = \frac{\mu^2\gamma^2}{2}\sim7\cdot10^{-79}\left(\frac{M}{10^6M_\odot}\right)^{-2},
\end{equation}
which is below the constrained values for astrophysical BH masses. The parameter $\gamma$ in the sigma models is related to the dimensionful energy scale of the scalar manifold $E$ as 
\begin{equation}
    |\gamma|=\frac{M_P}{\sqrt{16\pi}E},
\end{equation}
with $M_P$ the Planck mass. It is natural to consider $E\ll M_P$, so that $\gamma$ can be large.

This work is a first step in understanding the phenomenology of matter fields from sigma models in an astrophysical scenario. Further investigations are, however, needed to first of all explore the theory parameter space, to gain a better understanding of the dependence of GW observables on the matter sector properties. Similarly, we leave for future work to assess the detectability of such environmental effects. For instance, the dephasing of the signal induced by matter distributions could lead to biases in parameter estimation when matched to vacuum GR templates \cite{Roy:2024rhe, DeLuca:2025bph}.


\vspace{0.4cm}
\begin{acknowledgments}   
\textbf{\textit{Acknowledgments.}}
 We thank Katy Clough, Josu Aurrekoetxea, Pablo Cano and Geoffrey Compère for useful comments on the draft. We also thank Sam Brady and Miguel Bezares for useful discussions. We thank the entire \texttt{GRTL} Collaboration\footnote{\texttt{www.grtlcollaboration.org}} for their support and code development work. LM is a PhD fellow at the Research Foundation - Flanders (FWO grant 1186024N). LM acknowledges funding from ESA Prodex project ’LISA EMRI/IMRAC waveform modelling’ PEA 4000131558, which financed earlier stages of this research. LAS is partly funded by IBOF/21/084. The resources and services used in this work were provided by the VSC (Flemish Supercomputer Center), funded by the Research Foundation - Flanders (FWO) and the Flemish Government. This work also used the DiRAC@Durham facility managed by the Institute for Computational Cosmology on behalf of the STFC DiRAC HPC Facility (www.dirac.ac.uk). The equipment was funded by BEIS capital funding via STFC capital grants ST/P002293/1, ST/R002371/1 and ST/S002502/1, Durham University and STFC operations grant ST/R000832/1. DiRAC is part of the National e-Infrastructure.
\end{acknowledgments}

\section*{Code availability}
We provide publicly the extensions to \texttt{GRChombo} and \texttt{GRTresna} that we encoded to simulate scalar fields from sigma models in the \href{https://github.com/ludomachet/Scalar_Field_from_Sigma_Models}{GitHub directory}
\cite{Machet:2025code}.

\appendix
\section{Matter equations in the ADM decomposition}
\label{sec:3+1}

In the $3+1$ decomposition, the Einstein-Klein-Gordon equation \eqref{eq:EKG} reduces to two first-order equations

\begin{align}
 &\Pi=\frac{1}{\alpha}(\partial_t\Phi-\beta^i\partial_i\Phi),\label{eq:EOM1}\\
 &\partial_t\Pi=\beta^i\partial_i\Pi+\alpha\bigg(K\,\Pi+D_iD^i\Phi\nonumber\\
 &\quad\quad-\frac{\gamma^2\bar\Phi}{2\left(1-\frac{\gamma^2}{4}|\Phi|^2\right)}(\Pi^2-D_i\Phi\,D^i\Phi)\nonumber\\
 &\quad\quad-2\Phi\left(1-\frac{\gamma^2}{4}|\Phi|^2\right)^2U'(|\Phi|^2)\bigg)\label{eq:EOM2}.
\end{align}

Two additional equations emerge for the complex conjugates $(\bar\Phi,\bar\Pi)$, which are considered as independent variables, sending $\Phi\rightarrow\bar\Phi, \,\Pi\rightarrow\bar\Pi$ in equations \eqref{eq:EOM1} and \eqref{eq:EOM2}.

After decomposing the Einstein field equations in the $(3+1)$ formalism, the stress-energy tensor components associated to the scalar field $\Phi$ read

\begin{align}
    &\rho = \frac{1}{2\left(1-\frac{\gamma^2}{4}|\Phi|^2\right)^2}(D_i\bar{\Phi} \,D^i\Phi+\Pi \, \bar\Pi)+U(|\Phi|^2),\\
    &S_i=\frac{1}{2\left(1-\frac{\gamma^2}{4}|\Phi|^2\right)^2} (\bar\Pi \,D_i\Phi+ \Pi\, D_i\bar\Phi),\\
    &S_{ij}=\frac{1}{2\left(1-\frac{\gamma^2}{4}|\Phi|^2\right)^2}\big(\gamma_{ij}\,(\Pi \, \bar\Pi-D_i\bar{\Phi} \,D^i\Phi)\nonumber\\&\quad\quad+D_i\bar{\Phi}D_j\,\Phi+D_i\Phi\,D_j\bar{\Phi}\big)-\gamma_{ij}\,U(|\Phi|^2).
\end{align}

\section{Convergence}
\label{sec:convergence}

\begin{figure}[htbp]
	\centering
	\includegraphics[width=0.5\textwidth]{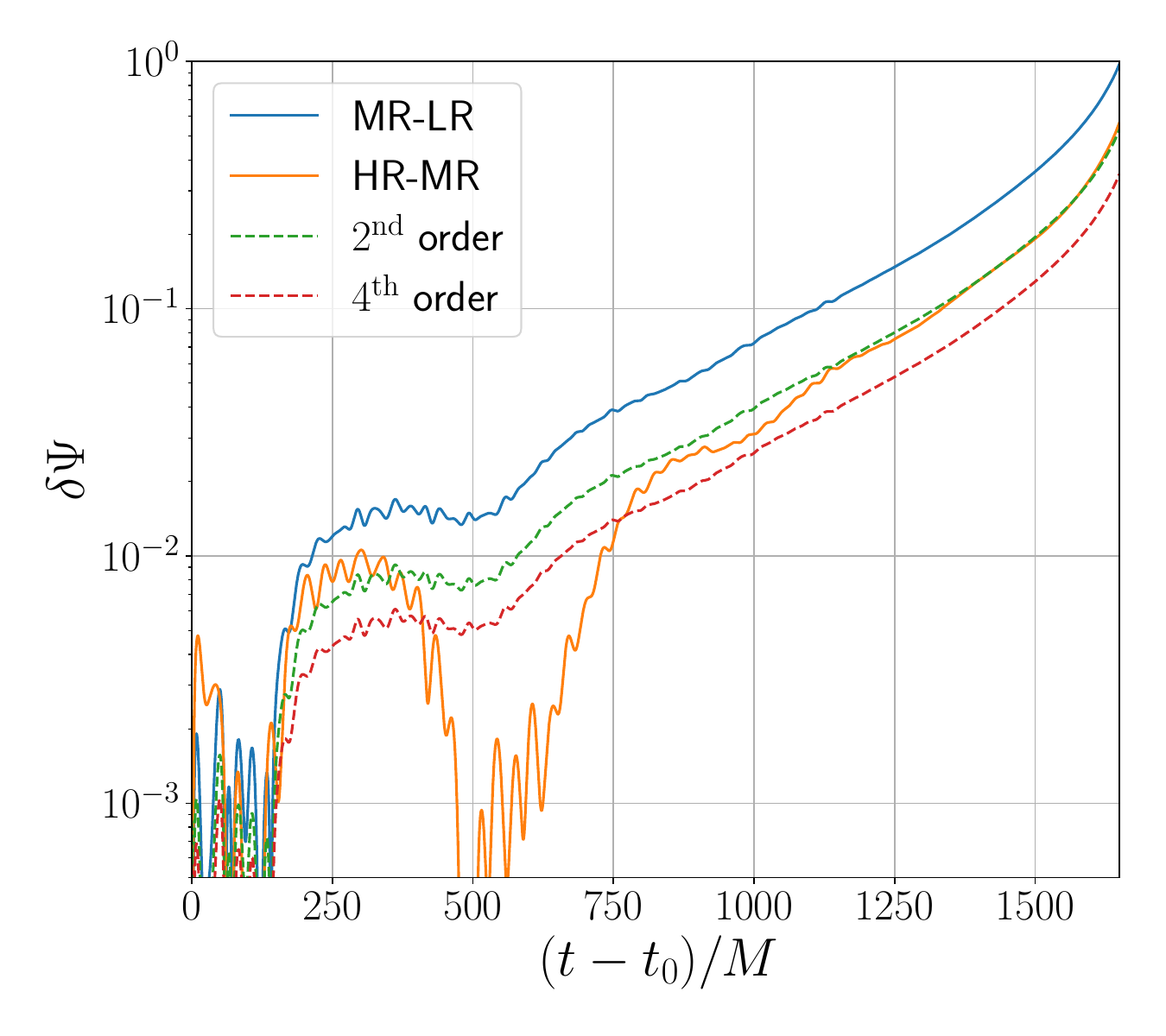}
	\caption{Difference in accumulated phase in between high resolution (HR), mid resolution (MR) and low resolution (LR) simulations of a BBH with scalar field mass $\mu M=0.3$ and $\gamma^2=10^5$.}
	\label{fig:convergence_test_GRChombo}			
\end{figure}

\begin{figure}[htbp]
	\centering
	\includegraphics[width=0.5\textwidth]{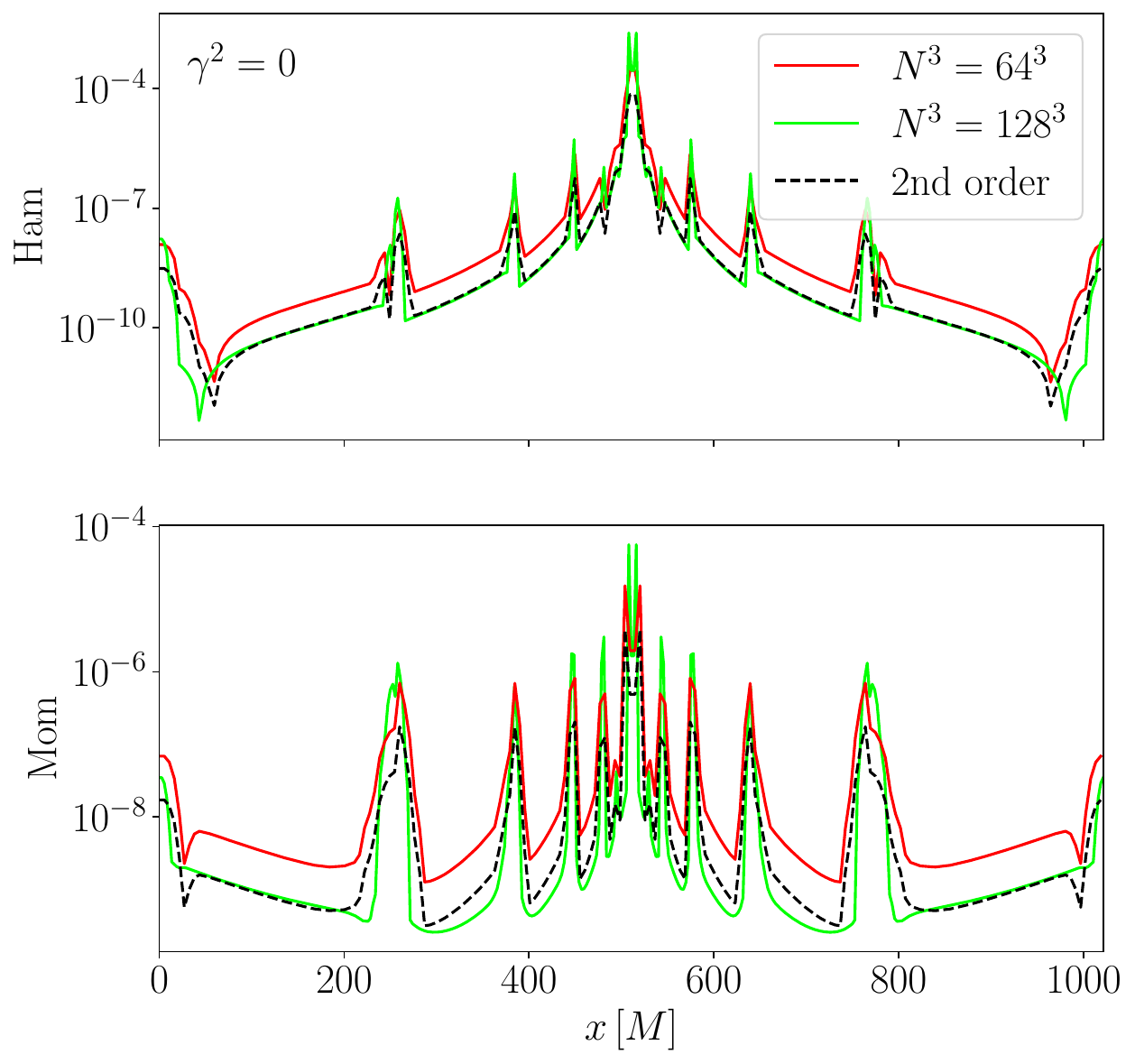}
	\caption{Difference across resolutions for the Hamiltonian and momentum constraints in space domain for the initial data obtained with \texttt{GRTresna}.}
	\label{fig:convergence_test_GRTresna}			
\end{figure}

In this section, we present the convergence test for the orbital phase $\Psi$ defined in this paper for the BBH system with scalar field mass $\mu M=0.3$ and $\gamma^2=10^5$. We show in Figure \ref{fig:convergence_test_GRChombo} its difference across resolutions for the simulations that we ran for this case, namely with a number of cells on the coarsest level of $N=128$, $N=160$ and $N=192$, respectively for low (LR), medium (MR) and high (HR) resolutions respectively. It demonstrates that the convergence order during the inspiral and merger of $\Psi$ is around two, with a mild overconvergence during a brief part of the inspiral which was also observed in the detailed study carried out in \cite{Radia:2021smk}. This is consistent with the order of convergence of our initial data solver \texttt{GRTresna}, which we show in Figure \ref{fig:convergence_test_GRTresna}. The
results of the convergence analysis presented here indicate that our simulations are stable and in
the convergent regime.

\bibliographystyle{apsrev4-1} 
\bibliography{AxiDIlaton_NR}

\end{document}
%